# Imaging mechanism and contrast separation in low-voltage scanning electron microscopy imaging of carbon nanotube arrays on SiO$_2$/Si substrate

Boxiang Zhang [a,b], Zhiyong Zhang [c,d], Chuanhong Jin [a,b,d*]

[a] State Key Laboratory for Silicon and Advanced Semiconductor Materials, School of Materials Science and Engineering, Zhejiang University, Hangzhou, Zhejiang 310027, China

[b] Jihua Laboratory, Foshan, Guangdong 528200, China

[c] Key Laboratory for the Physics and Chemistry of Nanodevices and Center for Carbon-based Electronics, School of Electronics, Peking University, Beijing 100871, China

[d] Hunan Institute of Advanced Sensing and Information Technology, Xiangtan University, Xiangtan, Hunan 411105, China

*Corresponding author. E-mail: chhjin@zju.edu.cn

## Abstract

Polymer-sorted high-density carbon nanotube (CNT) arrays have shown great potential to extend the silicon-based Moore's law. Imaging the CNT arrays on insulators like SiO$_2$/Si using low-voltage scanning electron microscopy (LVSEM) to acquire array information like the alignment, density, and distribution of residual polymers is necessary. Such a task remains challenging due to the nanoscale CNT body (1-2 nm in diameter), nanoscale tube-to-tube separation (1-10 nm), the broadening of the apparent diameter, and the complex image contrast caused by the insulating substrate and polymer residues. In this study, the imaging mechanism for this system is investigated. Two methods are developed to separate the three dominant contrasts, *i.e.* topographic contrast, charge contrast, and material contrast, by selecting the take-off angle and energy of the emitted electrons as enabled by changing the working distance or the deceleration voltage. The contrast formation and separation mechanism is further confirmed by the dynamic contrast evolution due to the electron-beam-induced deposition of amorphous carbon. The contrast separation method is further applied to an individual CNT, reducing its apparent diameter from 36 nm to 6 nm. This result hints at the potential for LVSEM to count the density exceeding 150 CNTs/μm of CNT arrays. Finally, a comparative study of LVSEM and transmission electron microscopy confirms the failure of LVSEM to resolve CNTs in a bundle. The results suggest that the density of CNT arrays may be underestimated in reported SEM data. The proposed method can serve as a useful tool for further study and application of arrayed CNTs.

1. **Introduction**

Semiconducting single-walled carbon nanotubes (CNTs) have exhibited great potential as the building block for next-generation high-performance and energy-efficient field effect transistors (FETs) and integrated circuits (ICs) due to their unique structure and excellent properties, such as the ultrahigh carrier mobility, atomically thin body and immunity to short channel effect. Although early successes on FETs built on individual CNTs (CNTFET) has already displayed promising application since the projected performance may go beyond the silicon technology [1–3], the real potential of CNTFETs has been demonstrated until very recently by a few research groups [4–8] on the polymer-sorted, high-density (>100 CNTs/μm) and high-purity (up to 99.9999%) semiconducting CNTs arrays.

Imaging the arrayed CNTs on insulating substrates and acquiring array information such as the density, alignment, and defects including the unavoidable polymer residues then becomes necessary for the fabrication, inspection, and further development of CNT array wafers and devices. Scanning electron microscope (SEM) is expected to play a vital role for this purpose owing to its high spatial resolution, the capability of automated wafer-scale characterization, and the compatibility of being integrated into the production lines. This has been demonstrated in mainstream silicon technology, where SEMs are widely used for critical dimension measurement, inspection, and review of fabrication processes and product devices.

Low-voltage (LV) SEM operating the microscope at low acceleration voltages (< 5 kV and typically below 2 kV) possesses a number of advantages over traditional high-voltage SEM (5 kV - 30 kV). The surface sensitivity and signal-to-noise ratio are all improved in LVSEM due to the reduced electron-beam interaction volume [9], and the increased ratio of the feature size to the electron-sample interaction volume increases the topographic contrast, which can hardly be obtained when high acceleration voltages are used [10]. Historically, LVSEM used to suffer from the limited obtainable beam size caused mainly by chromatic aberration at low voltages, while this issue has been addressed in large part with the development of modern electron-optical-column design such as the introduction of the magnetic-electrostatic compound lens and beam deceleration technology [11]. In modern SEMs, a high beam energy is maintained throughout the entire column, decreasing chromatic aberration and ensuring a sufficiently small beam [11–14] that lands on the sample surface. Recent advances in electron detector technology have further enabled differentiating signal electrons in terms of the take-off angle [15,16] and kinetic energy [17,18] using different detectors. Hence, the contribution of different contrast components such as topographic, material, and charge contrasts to the final LVSEM image becomes separable since signal electrons in a specific angular and energy range predominantly contribute to a specific kind of contrast.

SEM, particularly LVSEM, has been widely used for characterizing low-density CNTs such as individual CNTs [19–27]. chemical vapor deposited (CVD) CNT arrays with a tube-to-tube separation (pitch) > 100 nm [28–30]. and network CNT films [19,22,31–34]. The imaging mechanism can be divided into two main categories, which are charge contrast, first proposed by Brintlinger *et al.* [21] and electron-beam-induced-current (EBIC), first proposed by Homma *et al.* [31] Charge contrast attributes the image contrast to the local potential difference between the CNTs and the substrate induced by the electron bombardment, whereas EBIC attributes the image contrast to the supply of electrons from the CNTs to the substrate and such electron supply reduces the substrate charging under electron-beam irradiation and helps to maintain its high initial electron emission yield.

However, two major challenges arise when applying LVSEM on the aforementioned polymer-sorted high-density CNT arrays loaded on an insulating substrate like $SiO_2$/Si. The first issue is stemmed from the well-known insulator-induced charging effect [21,22,31] (the associated imaging contrast is denoted as charge contrast), which makes the apparent diameter of a single CNT measured from an LVSEM image typically broaden to over 30 nm and sometimes even ~100 nm in contrast to its actual diameter of only 1-2 nm [21,22,31]. Such a diameter broadening causes the neighboring CNTs (with a pitch typically < 30 nm) in a high-density CNT array hard to distinguish under LVSEM. The second issue is related to the insulating polymer residues. The thickness and spatial distribution of residual polymers are processing-dependent, hard to control [35–37], and thus rather inhomogeneously distributed across the entire wafer at the present stage, complicating the LVSEM image formation and interpretation. A prerequisite to addressing these challenges is the elucidation of the LVSEM contrast mechanism, which is a necessity for obtaining reliable information and avoiding misinterpretation of LVSEM images of the insulator-supported high-density CNT arrays.

This study aims to separate three major contrast components: topographic, material, and charge contrasts of polymer-sorted CNT arrays, interpret the LVSEM images of polymer-sorted

CNT arrays, and further point out the possible limitations. Two methods are proposed and demonstrated for contrast separation by selecting the angular and energy range of emitted electrons using two different mainstream LVSEM instruments. The application and limitation of the contrast separation method in LVSEM of CNTs are also presented and discussed.

## 2. Materials and methods

Polymer-sorted CNT arrays on $SiO_2$/Si wafers were purchased from Suzhou Carbon Technology Co. Ltd with the detailed preparation process reported previously [5]. Briefly, arc-discharge SWCNTs were sonicated, dispersed, and then centrifugated in toluene with the assistance of polymer (poly[9-(1-octylnonyl)-9Hcarbazole-2,7-diyl], PCzs) wrapping. The sorted CNTs with a semiconducting tube purity >99.9999% were then assembled into arrays on thermally grown $SiO_2$ (285 nm)/Si (525 μm) substrate through a dimension-limit self-alignment (DLSA) procedure, followed by a cleaning process to remove residual polymers partly. The low-density CNT array (<10 CNTs/μm, free of any polymer) was chemical vapor deposited on quartz and then transferred to $SiO_2$/Si substrate following the reported procedure [46].

For general purpose, two state-of-the-art LVSEM systems capable of the selection of signal electrons with different energies and angles were employed with one from Carl Zeiss Inc. and the other one from Hitachi Inc. The key components for Zeiss LVSEM (Model Gemini 500/300) include a Schottky gun, a magnetic-electrostatic hybrid lens system, an 8 kV beam booster for in-column beam acc/deceleration and an Inlens detector. The Hitachi LVSEM (Model Regulus 8230) comprises a cold field emission gun, an immersion magnetic lens system, and an ExB velocity filter. In the Hitachi LVSEM, two detectors in different locations are used. Both the Upper detector and the Top detector in the Regulus 8230 collect the secondary electrons generated following collision and amplification with the convertor electrode by the HE electrons. LEs are deflected and collected by the Upper detector because of the ExB filter. SEM imaging parameters are listed in Table S1. To reduce the influence of carbon contamination and accumulated charging, we selected a fresh area for each image unless stated otherwise. The sample for Figure 7 was left in the air for two weeks in advance to accelerate the electron-beam-induced amorphous carbon deposition. The contrast and intensity of the detectors were adjusted for maximum visibility.

The cross-section sample of the CNT array on SiO$_2$/Si substrate was prepared using a FEI Quanta 3D Nanolab Dual Beam (SEM/FIB) system. Annular dark-field and bright-field scanning transmission electron microscopy (ADF-STEM and BF-STEM) imaging was carried out using a FEI Titan ChemiSTEM, which was equipped with a probe-side spherical aberration corrector and operated at an accelerating voltage of 200 kV.

## 3. Results and discussion

### 3.1. Theoretical analysis of LVSEM imaging mechanism of CNT arrays

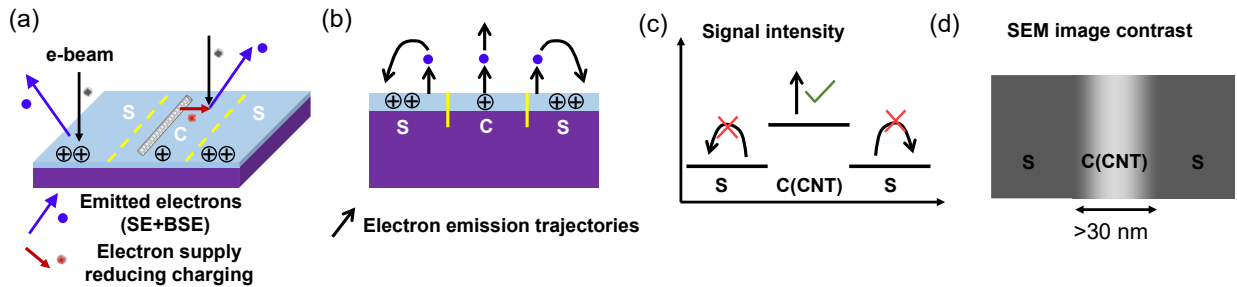

Figure 1 Schematics illustrating the charge contrast of a single CNT on the SiO$_2$/Si substrate. (a) The positively charged substrate because the emitted electrons are more than the incident electrons. The CNT provides electron supply to the surrounding substrate in region C (C for CNT), so region C is charged less positively than region S (S for substrate). (b) Electron emission trajectories in different regions. Region C, with a width of >30 nm, emits more electrons to the detector. (c-d) The schematic drawing of the signal intensity and SEM image contrast of the sample.

To explain why LVSEM imaging is not directly applicable for high-density CNT arrays, we first discuss the apparent diameter broadening phenomenon of CNTs, and the associated mechanism [21,22,31] is illustrated in Figure 1. When imaging a single CNT on the $SiO_2$/Si substrate, $SiO_2$ far from the CNT (Region S in Figure 1) are positively charged since the emitted electrons (whether they are secondary electrons (SE) or backscattered electrons (BSE)) are more than the incident electrons [9,38] (also illustrated in Figure S1a). In contrast, the CNT and its proximate $SiO_2$ (Region C in Figure 1) are relatively less positively charged since the CNT, being an electron reservoir, can supply electrons to nearby positively charged $SiO_2$, reducing the charging effect therein (also illustrated in Figure S1b). As a result, electrons emitted from region S are more likely attracted back rather than reaching the detector, making the signal intensity of region S lower [38] than that of region C (as shown in Figure 1c). Further considering the electron supply length between the CNT and proximate $SiO_2$ is typically greater than 15 nm [22,31], CNT thus appears brighter and wider (> 30 nm) in SEM images, as shown in Figure 1d.

However, such a charge contrast for individual or low-density CNTs on insulating substrate under LVSEM will become invisible for high-density CNT arrays, and the associated principle is illustrated in Figure 2. When the pitch decreases to about 30 nm (corresponding to a density of 33 CNTs/μm), the electron-supply regions bounding to each of the neighboring CNTs overlap with each other, making the neighboring CNTs not discriminable. As a result, the charge contrast is lost in LVSEM images of high-density CNT arrays.

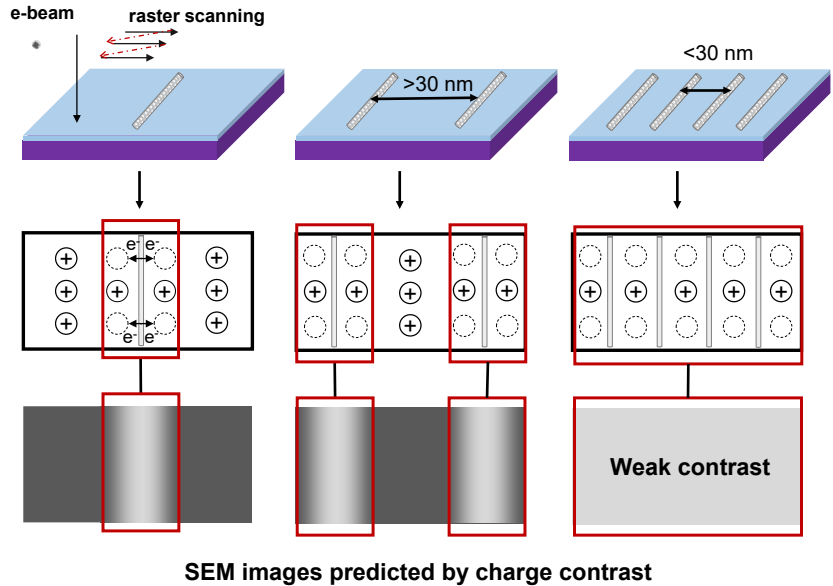

**Figure 2** Schematics illustrating the charge contrast of a single CNT, a low-density CNT array (pitch> 30 nm), and a high-density CNT array (pitch< 30 nm) on the SiO$_2$/Si substrates. The charge contrast is lost in the LVSEM image of high-density CNT arrays since the electron supply regions overlap with each other.

As illustrated in Figure 3, polymer-sorted CNT arrays loaded on the insulating substrate consist of high-density aligned CNTs and unavoidable residual polymers, thus three kinds of contrast components should be considered for LVSEM imaging: material, charge, and topographic contrasts. Actually, it is mainly the sizable polymer aggregations (a few nm to tens of nm) that alter the local material and charge contrast. In terms of material contrast, it is caused by the material-dependent electron emission yield difference between the polymer (such as PFO-BPy and PCz), CNT, and SiO$_2$, making them distinguishable. Due to the poor electrical conductivity of the insulating polymer residues, they can alter the charge distribution locally, making the lost charge contrast in high-density CNT arrays (shown in Figure 2) reappear again, as shown in Figure 3. The nanoscale height fluctuation of the CNTs modifies the electron-sample interaction and the electron emission, thus creating the topographic contrast. Polymers are assumed to be thin and bind to CNTs, and as a result, they make a negligible change to the array-like topographic contrast. This is reasonable since the as-prepared CNT arrays were cleaned for cycles, and the polymers that did not bind to CNTs were removed.

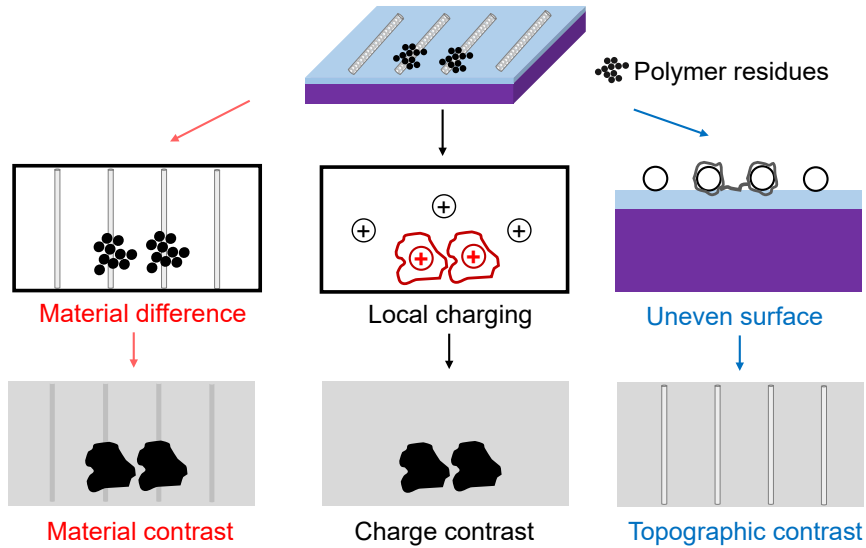

**Figure 3 Three contrast components for LVSEM of a polymer-sorted CNT array on SiO$_2$/Si. Note that the relative brightness between polymers, CNTs, and the substrate is established based on experimental results.**

In a real LVSEM image, these contrast components are usually entangled due to the detector's simultaneous reception of all types of signal electrons under certain operating parameters of the microscope. To obtain specific sample information, we should separate the three contrast components, which can be accomplished by separating the emitted electrons based on their kinetic energies and take-off angles (Figure 4). Note that we prefer to use the expression "energy and take-off angle" instead of "SE/BSE" to understand the subsequent contrast separation methods better. Comparing the energy distribution of the emitted electrons between the charge-balanced (red curve) and uncharged (black curve) cases, as schematically illustrated in Figure 4a, we can find that a few volts of surface potential resulting from charging is significant in altering the energy spectra, suppressing the emission of electrons, and achieving equilibrium between the incident and emitted electrons. In other words, charging mainly affects the emitted electrons in the energy range of 0-

10 eV [39]. The appearance of the large peaks indicated by the arrows in Figure 4a hints that most of the emitted electrons carry an energy of less than 10 eV and are thus sensitive to charging (meaning their trajectories are easily deflected due to the charging-induced electrical field), so the charge contrast is expected to command a dominant role, and may even supersede any apparent material and topographic contrasts in light of the abundance of sub-10 eV electrons. Emitted electrons with different take-off angles (the angle between the trajectory of emitted electron and the surface plane) also carry different information, as shown in Figure 4b, and they can be roughly divided into two major groups: low-angle (LA) and high-angle (HA) electrons. LA electrons predominantly carry topographic information since the emitted electrons prefer to leave from the lateral surface of CNTs [10]. In contrast, HA electrons carry more material information because the difference in the electron emission of varied materials becomes more significant at higher take-off angles [10,40] (see Figure S2 for more details). Effects of these two factors on the dominant contrast are summarized in Figure 4c, from which we can learn that for contrast separation, all we have to do is to find a way to receive emitted electrons with different take-off angles and energies.

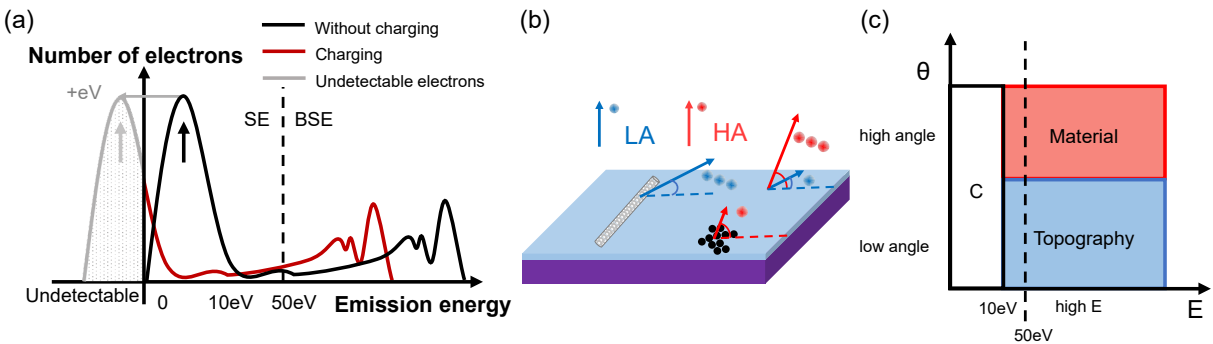

Figure 4 Schematics drawings of emitted electrons contributing to different contrasts. (a) Energy distribution of the emitted electrons showing how the charge contrast is dominated by low-energy electrons. (b) Low-angle (LA) electrons and high-angle (HA) electrons emitted from different regions showing that the topographic contrast is dominated by LA and the material contrast is dominated by HA. (c) The acceptance map illustrating the dominant contrast for electrons in certain take-off angle (θ) and energy (E) ranges. (C, Charge contrast)

### 3.2. Contrast separation by changing the working distance in a Zeiss GEMINI system

The first method is to change the working distance (WD, the distance between the sample and pole-piece) and thus electrons collected by the Inlens detector, as demonstrated on a Zeiss GEMINI system. This method has been reported in the characterization of steel surface [41,42], and the associated mechanism is illustrated in Figure 5a. As the WD increases, *i.e.* from 1 mm to 5 mm, the low take-off angle (LA) emitted electrons that mainly carry topographic information (recall the summarized mechanism in Figure 4) will fall out of the InLens detector, hence rarely collected, as illustrated by the blue trajectories in Figure 5a. In contrast, those high take-off angle (HA) emitted electrons which are rarely collected at WD = 1 mm, start to hit the detector as the WD increases (red trajectories in Figure 5a), enhancing the material contrast in the LVSEM images. The Zeiss GEMINI lens applies an upward electrical field force on emitted electrons [12,17,43], causing charged electrons with energies usually below 10 eV to deflect upwards easily, meaning that they are equivalent to emitting at a high take-off angle from the surface. Hence, the trajectories of charging electrons and HA electrons are similar, and charge contrast also increases as the WD increases. Actually, it is practically difficult to separate the material contrast and charge contrast because they are intrinsically entangled, and one solution is to use energy-filtered HA-BSE imaging. The angular and energy ranges of emitted electrons received by the Inlens detector at different WDs mentioned above are summarized in Figure S3.

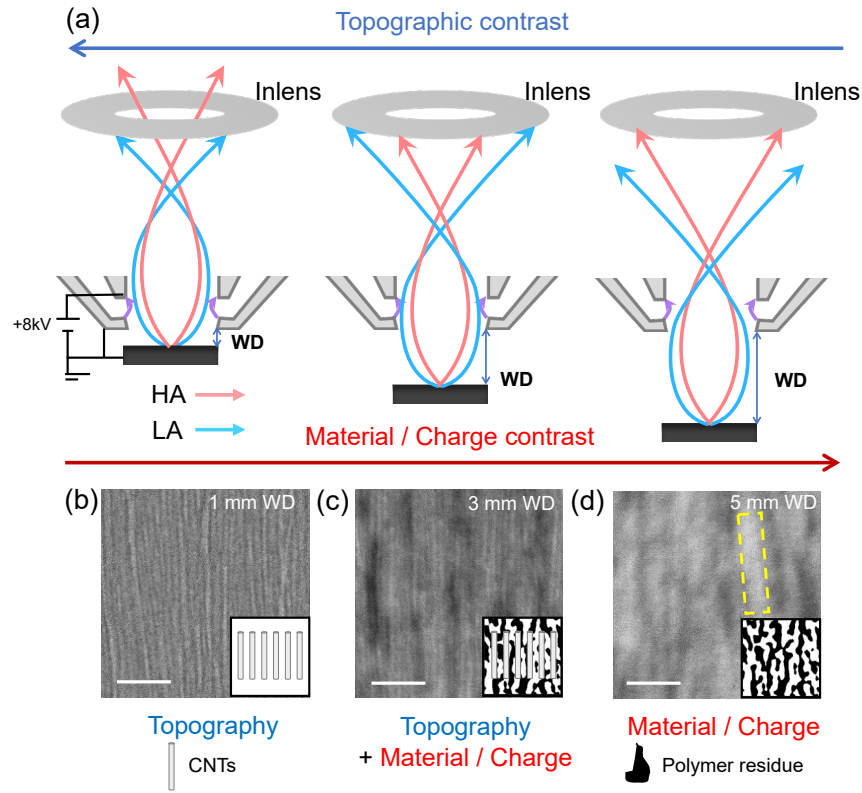

**Figure 5 Contrast separation by changing the WD. (a) Sketch drawing of the trajectories of LA and HA in the Zeiss GEMINI system as a function of WD. (b-d) LVSEM images of polymer-sorted CNT array obtained by the Inlens detector with WDs of (b) 1, (c) 3, and (d) 5 mm. The insets show the corresponding schematic illustrations of the SEM images. Scale bar: 100 nm**

Such an imaging mechanism analysis is well reproduced by the experimental results shown in Figure 5b-5d. At WD = 1 mm, the LVSEM image in Figure 5b mainly exhibits a feature of aligned rods. The measured width of these rods using the full width at half maximum (FWHM) method is in the range of 4 nm to 11 nm, much narrower than that of the isolated and low-density CNTs in LVSEM images [21,22,31] (typically 30-100 nm), so we suggest the rods-like features as the topographic contrast of CNTs. When the WD increases to 3 mm, a number of dark patches overlapping with the rods appear, as shown in Figure 5c. Further increasing the WD to 5 mm, the rods are barely visible, and the bright and dark patches become the main feature in Figure 5d, indicating the material and charge contrast caused by the polymer residues get dominant. The bright and dark patches can be determined respectively as the polymer-deficient regions and polymer-enriched regions. Evidence supporting this claim is given in Figure S4 where such bright patches (polymer-deficient) increase significantly in areas after the polymer-removal process.

One may argue that the bright patches in Figure 5d, marked by a yellow rectangle, might be the CNTs as they overall also show sort of aligned arrangement, and the dark patches could be assigned as the substrate regions, similar to previous studies on LVSEM of low-density CNTs [21,22,31]. However, this assignment does not hold valid in this case because the pitch of this high-density CNT array is smaller than 20 nm (see the STEM image in Figure 9c for atomic-scale evidence), so the whole substrate can get electron supply from CNTs, and the traditional charge contrast should disappear, as illustrated in Figure 2.

### 3.3. Contrast separation by changing the deceleration voltage in a Hitachi Regulus 8230

The second method we develop for contrast separation is to change the deceleration voltage ($V_{dec}$) and thus the electrons collected by the Upper and Top detectors, as demonstrated on a Hitachi Regulus 8230 system. The associated mechanism is illustrated in Figure 6. The ExB is an energy filter that does affect the incident electron beam but deflects the emitted electrons toward the Upper detector. When the $V_{dec}$ is set to zero, the red arrows in Figure 6 illustrate that all the low-energy (LE) electrons entering the ExB, whether they are HA or LA electrons, are collected by the Upper detector because the ExB has a strong deflection effect on the LE electrons [44,45]. Although some LA electrons with high-energy (HE) containing topographic contrast are also collected by the Upper detector, their contribution is quite limited since the number of sub-10 eV charging electrons is much larger therein. Therefore, the charge contrast, *i.e.* bright and dark patches caused by polymer residues, are mainly observed in Figure 6a. As the $V_{dec}$ increases from 0 to -300 V, the deceleration electric field generated for the primary beam acts as an acceleration field for electrons emitted from the sample, thus providing sufficient energy to the LE electrons to pass through the ExB filter, and meanwhile increasing the effective take-off angle of LE electrons. As a result, LE electrons that carry mainly charging information hit the convertor plate of the Top detector at a high $V_{dec}$ but can't hit any detector at a low $V_{dec}$, as illustrated by the black trajectories in Figure 6. In contrast, HE electrons are less sensitive to the external electric field caused by $V_{dec}$ or charging, so the LA-HE electrons that carry mainly topographic information always hit the Upper detector, irrespective of the varied $V_{dec}$. The topographic contrast will be more obvious considering the take-off angles of the electrons accepted by the Upper detector are slightly lower when the $V_{dec}$ is higher. The angular and energy ranges of emitted electrons received by the two detectors with different deceleration voltages mentioned above are summarized in Figure S5.

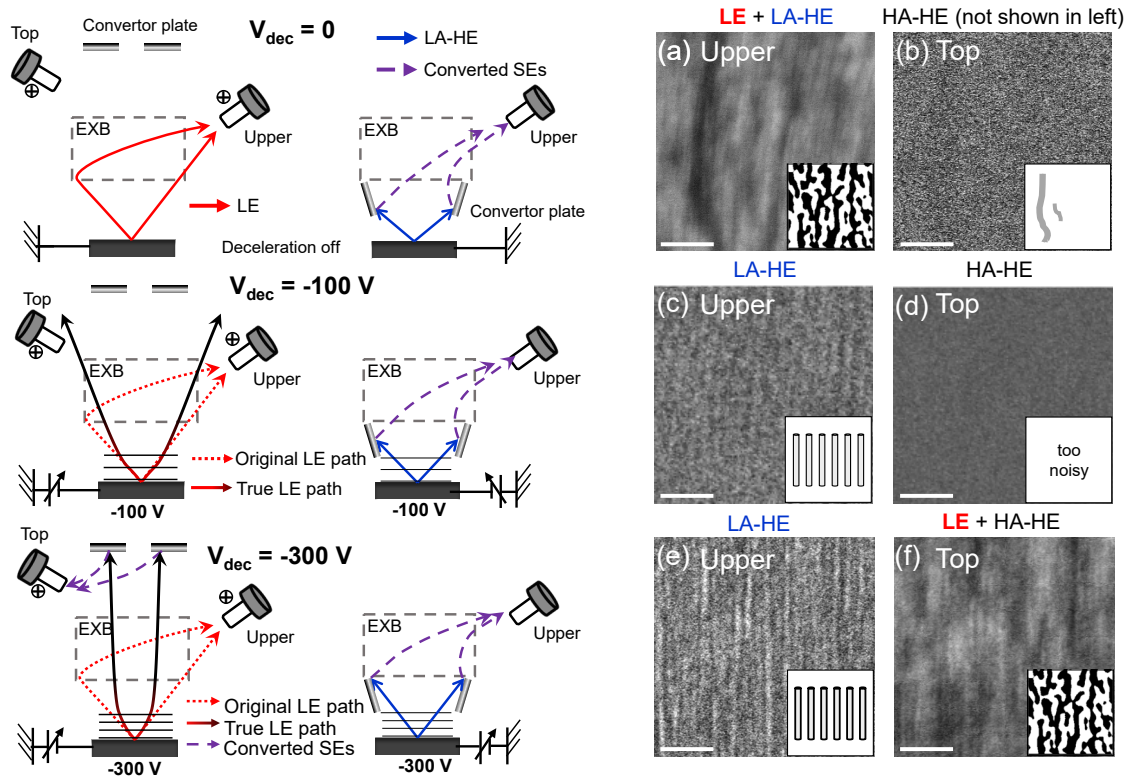

**Figure 6 Contrast separation by changing the deceleration voltage in the Hitachi Regulus8230.** LVSEM images of polymer-sorted CNT array obtained by the (a, c, e) Upper detector and (b, d, f) Top detector with a deceleration voltage of (a, b) 0, (c, d) -100 V, and (e, f) -300V. The schematic drawing illustrates the electron trajectories at different deceleration voltages. The red and dashed red arrows represent the trajectories of low-energy (LE) electrons with the deceleration mode off. The black arrows represent the trajectories of LE electrons with a deceleration voltage, and the blue arrows represent the trajectories of low-angle and high-energy electrons (LA-HE). Scale bar: 100 nm

Such an imaging mechanism analysis is justified by the experimental results shown in Figure 6a-f, where the LVSEM images are recorded simultaneously by the Upper (a, c, e) and Top detectors (b, d, f) at different $V_{dec}$. The bright and dark patches are observed in the Upper detector in Figure 6a at $V_{dec}$= 0 V and in the Top detector in Figure 6f at $V_{dec}$= -300 V, indicating the material and charge contrasts caused by the polymer residues are collected. The rods are faintly observed in the Upper detector at $V_{dec}$= -100 V in Figure 6c since the Upper detector collects LA-HE electrons that mainly carry topographic information. The rods become more clearly visible in Figure 6e since the Upper detector collects electrons with lower take-off angles (carry more topographic information) at $V_{dec}$= -300 V, thus enhancing the topographic contrast.

It should be noted that theoretically, the pure material information carried by HA-HE electrons can be completely separated by the Top detector without or with a low $V_{dec}$, as all the LA and LE electrons carrying topographic and charging information are deflected into the Upper detector and filtered out under these circumstances. However, Figure 6b and 6d acquired by the Top detector show that this ideal situation is hard to establish experimentally due to the limited signal-to-noise ratio, where only a minor/trace of material contrast is noticeable. This is because HA-HE electrons are almost vertically reflected, and their amount is very limited.

### 3.4. Dynamic contrast evolution justifying the contrast formation and separation mechanism

The proposed contrast formation and separation mechanism for LVSEM of polymer-sorted CNT arrays can be further justified by the dynamic contrast evolution during the repeated scanning process. Figure 7 shows the SEM images recorded repeatedly from the same scan area in a Zeiss GEMINI system at a WD of 5 mm, during which an amorphous carbon film is established onto the CNT array due to the well-known electron-beam-induced deposition process. The major image feature evolves from the original bright and dark patches (corresponding to material/charge contrast induced by the polymer residues, 1st scan, Figure 7a) to the concurrence of some aligned rods (corresponding to the topographic contrast of CNTs) and patches (2nd scan, Figure 7b), and then to almost aligned rods only (3rd scan, Figure 7c).

Such a contrast evolution can be quantitatively analyzed by plotting the line profile of greyscale values along the long side of the rectangles from the same region (red: Figure 7a, black: 7b, and blue: 7c), and the results are shown in Figure 7d. For better comparison, all the greyscale curves are obtained under the same detector setting, smoothed, and normalized to (0,1). The three rods (yellow-arrowed in Figure 7b and 7c) are chosen as an indicator. As seen in Figure 7d, no peak in the red curve is observed at the position of the three rods, indicating they are invisible during the first scan. In contrast, three peaks at the rod position are recognized in the black (from Figure 7b) curve and more clearly recognized in the blue curve (from Figure 7c). If we define the topographic contrast as $|I_A-I_B|/(I_A+I_B)$ and charge/material contrast as $|I_A-I_C|/(I_A+I_C)$, where $I_A$, $I_B$, and $I_C$ stand for the greyscale values at position A, B, and C in Figure 7d, the numerical contrasts are thus determined as summarized in Figure 7e.

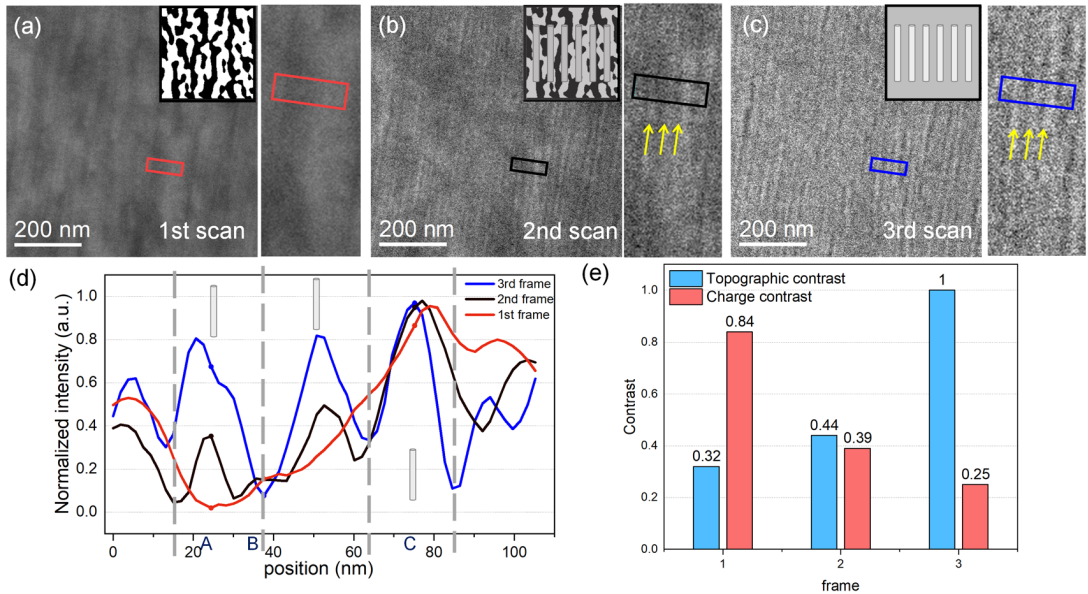

Figure 7 Contrast evolution due to the electron-beam-induced deposition in LVSEM images of polymer-sorted CNT array obtained by the Inlens detector at a WD of 5 mm during the repeated scan process. (a) The first scan image. (b) The second scan image. (c) The third scan image. (d) Smoothed line profiles along the long side of the rectangles in (a, b, c) and averaged along the short side. (e) Topographic and charge contrast values of the three images. Yellow arrows indicate the three rods in (b, c) corresponding to the three peaks in (d). (Zeiss GEMINI)

The origin of the contrast evolution is then discussed. During the electron beam rastering, a thin layer of amorphous carbon conformal to the CNTs is deposited onto the polymer-sorted CNT arrays due to the electron-beam-induced deposition. In contrast to the insulating polymers and $SiO_2$, this layer is more conductive and thus can mitigate the charging. It also reduces the compositional difference between the CNT (carbon), $SiO_2$, and polymer residues. As a result, the charge and material contrasts are progressively suppressed and even become almost invisible during the repeated scanning, making the topographic information show up or even dominant, as summarized in Figure 7e. Note that if we compare the results shown in Figure 7c with that in Figure 5c, they look similar overall, while the difference is that the topographic contrast of CNT arrays in Figure 7c is much more blurred than that in Figure 5c. The reason is that although the material and charge contrasts are suppressed in Figure 7c, the signal electrons are indeed HA electrons carrying less topographic information, thus appearing more blurred.

### 3.5. Contrast separation in low-density CVD-grown CNTs

To verify the feasibility of the proposed contrast separation methods in low-density CNTs, we apply them to LVSEM imaging of individual CVD-grown CNT as a representative. Figure 8a and 8b show the LVSEM images of the same CNT recorded simultaneously by the Top detector for Figure 8a and the Upper detector for Figure 8b at $V_{dec}$= -300 V. As seen, the apparent width of the rod has significantly decreased. Quantitatively, the FWHM measured from the integrated line profiles reduces from 36 nm (Top detector, collecting mainly LE charging electrons) to 6 nm (Upper detector, collecting LA-HE electrons mainly carrying topographic information). To our knowledge, a FWHM width as narrow as 6 nm of an isolated CNT on insulators using LVSEM is the best result so far, proving that the contrast separation mechanism we proposed also works for LVSEM imaging of low-density CNTs. The reason why topographic and material contrasts of low-density CNTs are not observed in previous studies [19–27] is that the number of low-energy charging electrons is large enough to drown out the other contrasts, as in Figure 5d.

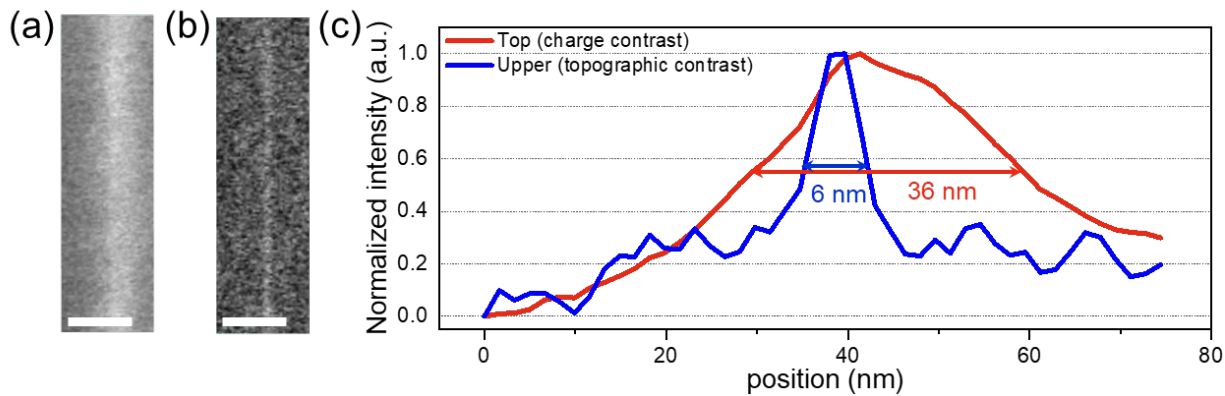

**Figure 8 LVSEM contrast separation of a CVD-grown CNT using a Hitachi Regulus 8230. (a-b) LVSEM images recorded by (a) the Top detector and (b) the Upper detector at a deceleration voltage of -300 V. (c) Corresponding line profiles averaged along the CNT. The width is taken as the FWHM after fitting with Gaussian curves.**

3.6. Limitations of the density determination of polymer-sorted CNT arrays using LVSEM

When using LVSEM for determining the density of CNT arrays, particularly for those with a pitch of < 6 nm, results from Figure 9 suggest a limitation exists: the density may not be counted accurately, even in the topographic contrast-dominant LVSEM images. For a cross verification, we also conduct the cross-sectional annual dark-field and bright-field scanning transmission electron microscope (ADF-STEM and BF-STEM) characterization of a chosen region of polymer-sorted high-density CNT arrays, which are first characterized via LVSEM (at 1 mm WD, the best condition for extracting topographic information), and the results are shown in Figure 9a-9c. As seen, the density counted from LVSEM is around 80 "CNTs"/μm (see Figure S6 for more details), but results from ADF-STEM and BF-STEM (Figure 9b and 9c) clearly show that each rod-like feature in the LVSEM image (Figure 9a) actually represents a CNT bundle, rather than a single CNT as misidentified by the LVSEM. In this specific area, the true density exceeds 240 CNTs/μm (80 bundles/μm), much denser than that determined by the LVSEM. This result also suggests additional cares are necessary when applying LVSEM on applications like counting the density of polymer-sorted CNT arrays.

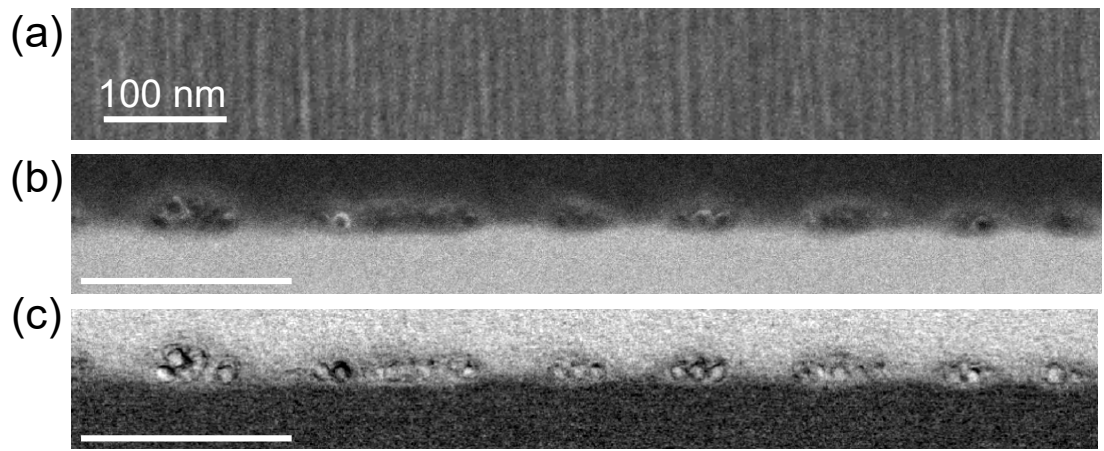

Figure 9 LVSEM and cross-section STEM images of a polymer-sorted CNT array. (a) LVSEM image obtained at a WD of 1 mm using a Zeiss GEMINI system. (b) ADF-STEM and (c) ABF-STEM images acquired simultaneously. The density of CNTs counted by SEM is approximately the same as the density of bundles counted by STEM.

## 4. Conclusions

In summary, the LVSEM imaging mechanism of polymer-sorted CNT arrays on $SiO_2$/Si substrates has been investigated. We analyze the origin of the three dominant contrasts: the topographic contrast, charge contrast, and material contrast in this system's LVSEM images. The topographic contrast is determined by the nanoscale height fluctuations caused by CNT arrays. The charge contrast is attributed to the positive charging of the polymer with respect to the CNTs and the substrate. The material contrast originates from the material-dependent electron emission yield between the polymer, substrate, and CNTs. Therefore, the material and charge contrasts reveal the distribution of polymers, and the topographic contrast reveals the alignment of CNTs. These three contrasts dominate in different angular and energy ranges, so contrasts can be separated if electrons with different take-off angles and energies are collected separately. We demonstrate two contrast separation methods in SEMs made by different manufacturers. The contrast mechanism is further justified by the dynamic contrast variation of LVSEM images due to the electron-beam-induced deposition of an amorphous carbon film. The contrast separation method is applied to the low-density CNT LVSEM imaging, narrowing the apparent diameter of the broadened CNT. The SEM images combined with the ADF-STEM and BF-STEM images indicate that SEM cannot distinguish between a small CNT bundle and a single CNT, even if the topographic contrast is mainly extracted. Therefore, the density of CNT arrays with densities can be under-counted, and bundling must always be considered when the LVSEM images of CNTs are analyzed. These results indicate that LVSEM has the potential to count the density of polymer-sorted CNT arrays with a density exceeding 150 CNTs/μm (corresponding to a pitch of 6 nm), but developments in LVSEM are still urgently needed for further applications in imaging denser CNT arrays.

## Acknowledgments

This work was financially supported by the National Key R&D Program of China (Grant No. 2022YFB4401602), the National Natural Science Foundation of China (Grant Nos. 51761165024 and 61721005), the Zhejiang Provincial Natural Science Foundation (Grant No. LD19E020002). This work was also supported by the Basic and Applied Basic Research Major Programme of Guangdong Province, China (Grant No. 2021B0301030003) and Jihua Laboratory (Project No. X210141TL210). The authors acknowledged the access to Zeiss Gemini 300 and Hitachi Regulus 8230 SEMs at Hangzhou Bay Institute (HZBI), which is affiliated to Ningbo Institute of Materials Technology and Engineering (NIMTE) under the Chinese Academy of Sciences (CAS), and to the Zeiss Gemini 500 SEM located at Westlake University.

**Supporting Information Available:**

Figures S1−S6, which illustrates the variation of electron emission yield (SE+BSE) with the incident beam energy; the contrast mainly contributed by different emission angles; the acceptance map of the Inlens detector at a WD of (a) 1 mm, (b) 3 mm, and (c) 5 mm; investigation of the black and white patches in the material/charge contrast dominant LVSEM images; the acceptance map of the Upper and Top detectors at a $V_{dec}$ of (a) 0, (b) -100 V, and (c) -300 V. Table 1, which summarizes the imaging parameters of the two LVSEM instruments.

**Supporting Information for**

**Imaging mechanism and contrast separation in low-voltage scanning electron microscopy imaging of carbon nanotube arrays on insulating substrate**

*Boxiang Zhang [a,b], Zhiyong Zhang [c,d], Chuanhong Jin [a,b,d]*

[a] State Key Laboratory for Silicon and Advanced Semiconductor Materials, School of Materials Science and Engineering, Zhejiang University, Hangzhou, Zhejiang 310027, China

[b] Jihua Laboratory, Foshan, Guangdong 528200, China

[c] Key Laboratory for the Physics and Chemistry of Nanodevices and Center for Carbon-based Electronics, School of Electronics, Peking University, Beijing 100871, China

[d] Hunan Institute of Advanced Sensing and Information Technology, Xiangtan University, Xiangtan, Hunan 411105, China

*Corresponding author. chhjin@zju.edu.cn

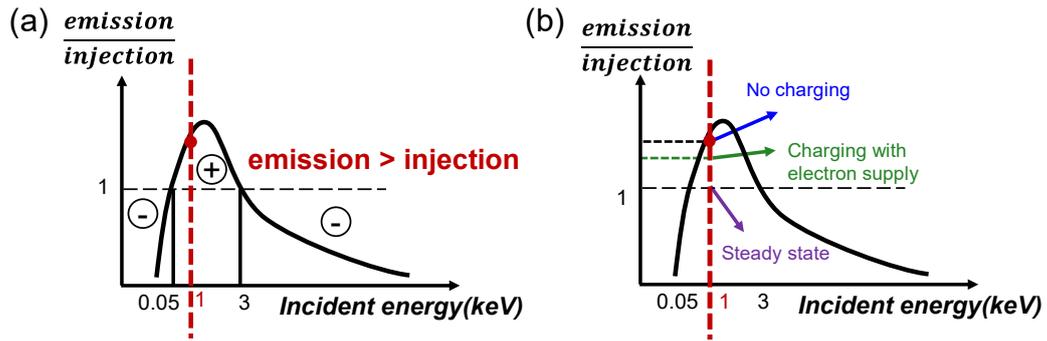

**Figure S1** Schematic drawing of the variation of electron emission yield (SE+BSE) with the incident beam energy. (a) Emission yield of $SiO_2$ at different incident energies before charging. (b) Emission yield of $SiO_2$ considering charging and electron supply of CNTs.

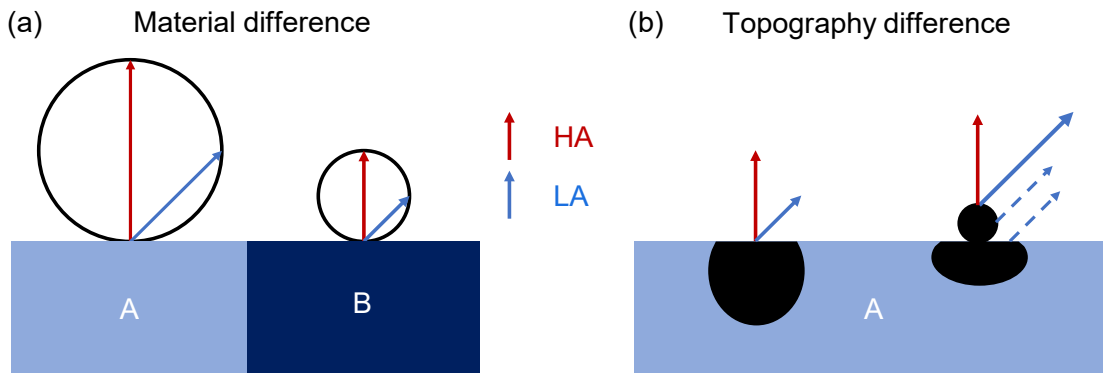

**Figure S2** Schematics of the contrast mainly contributed by different emission angles. (a) The material contrast dominated by the high-angle electrons. (b) The topographic contrast dominated by the low-angle electrons.

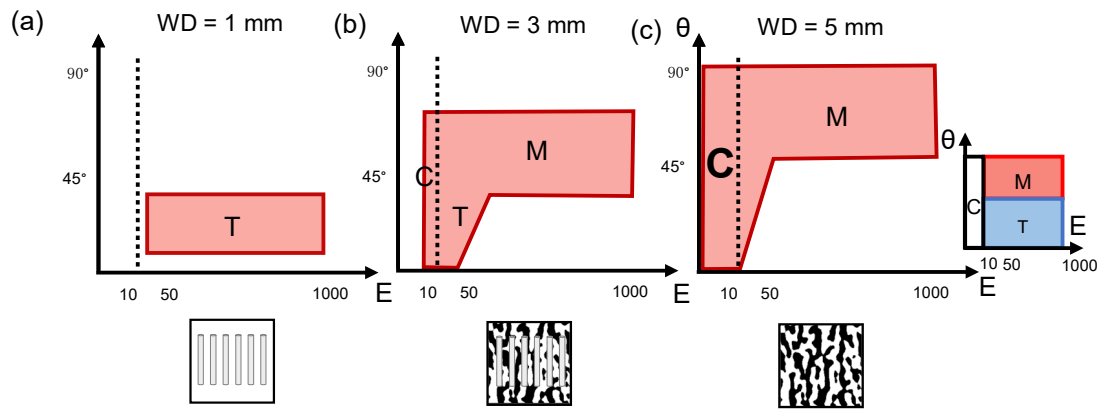

**Figure S3** The acceptance map of the Inlens detector at a WD of (a) 1 mm, (b) 3 mm, and (c) 5 mm.

The topographic contrast decreases as the WD increases, and the material and charge contrasts increase as the WD increases.

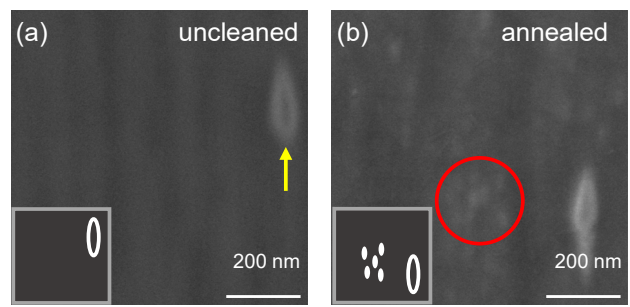

**Figure S4** SEM images of an uncleaned polymer-sorted CNT array. (a) Before and (b) after annealing in Ar gas at 350 °C for 1 hour. A WD of 5 mm is used to emphasize the material difference. The yellow arrow indicates the contrast referece, i.e. bright contrast of the bare substrate near the edge of the CNT arrays. White spots circled in red indicate the partial removal of the covered polymers.

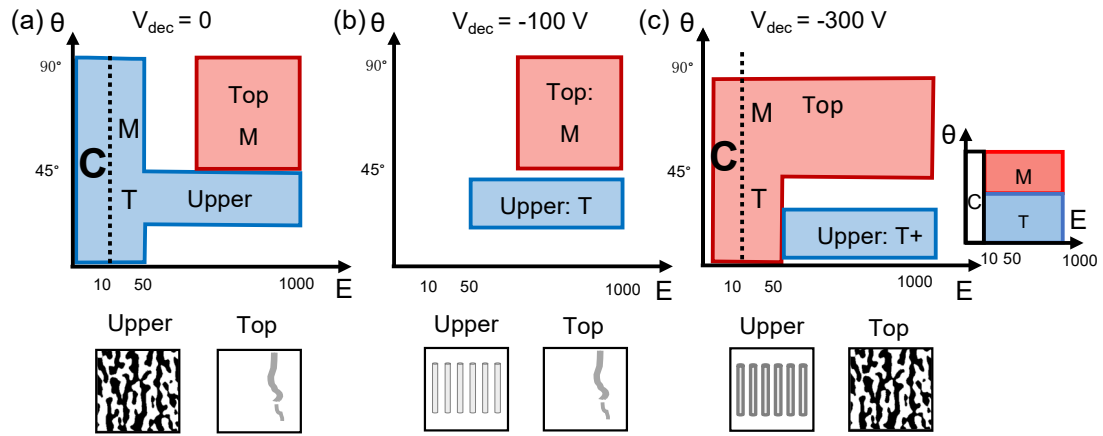

**Figure S5** The acceptance map of the Upper and Top detectors at a $V_{dec}$ of (a) 0, (b) -100 V, and (c) -300 V. The topographic contrast in the Upper detector increases as the $V_{dec}$ increases, and the material and charge contrasts transfer from the Upper detector to the Top detector as the $V_{dec}$ increases.

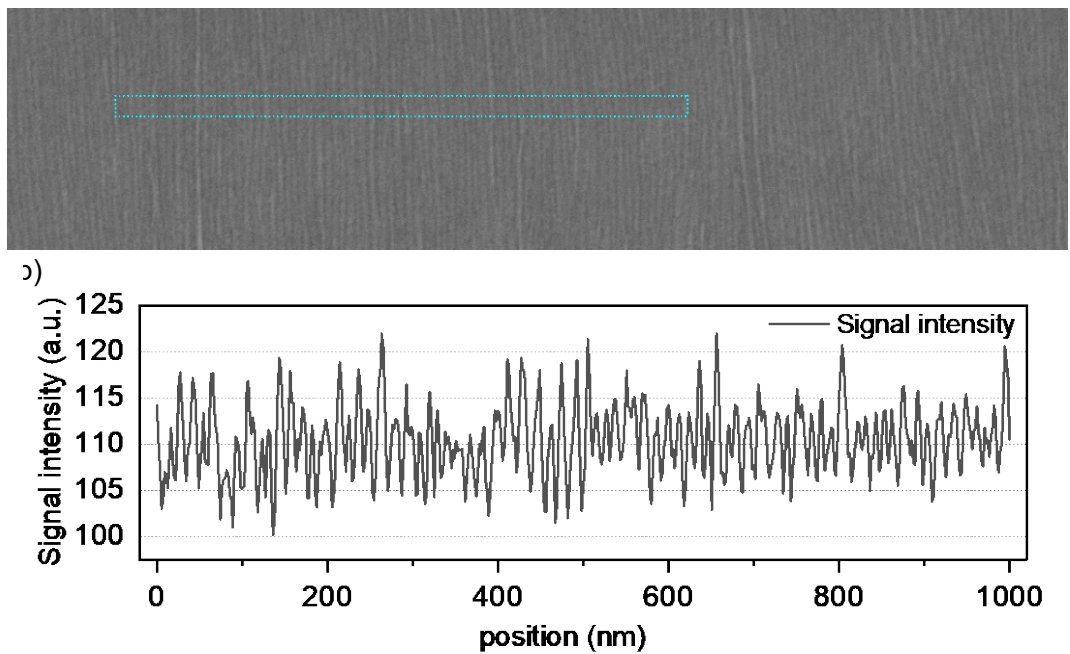

**Figure S6** Statistics of the number of CNTs in 1 μm in the typical region. (a) A LVSEM image obtained by the Inlens detector at a WD of 1 mm. (b) Corresponding line profile indicating 78 peaks in 1 μm, and the whole region looks similar. The region was characterized by STEM after SEM imaging.

**Table 1** Imaging parameters of the two SEM instruments. The beam current is measured by a Faraday cup stemmed on the sample stage in the Gemini 500/300, and measured by a Faraday cup and a picoampere meter without the stage bias in the Regulus 8230.

|  | Landing energy(keV) | Beam current(pA) | Pixel size(nm) | Dwell time(ns) | Line averaging count | Magnification | WD (mm) |
|---|---|---|---|---|---|---|---|
| Gemini 500/300 | 1 | 29.4/25.6 | 1.86 | 400 (Scan speed 3) | 60 | 60K | 1/3/5 |
| Regulus 8230 | 1 | 50.0 | 1.65 | 194 (CSS40 Capture) | 128 | 60K | 2 |